\documentstyle{jobp}

\begin{opening}
\title{CLIFFORD ALGEBRA DERIVATION of the
CHARACTERISTIC HYPERSURFACES of MAXWELL'S EQUATIONS}
\author{William M. Pezzaglia Jr.}
\institute{Department of Physics and Astronomy\\San Francisco State
University\\1600 Holloway Avenue\\San Francisco, California 94132\\U.S.A.}
\date{(Received: October 1992)}
\end{opening}

\runningtitle{Preprint \# SFSU-Th-92-05}

\begin{document}
\begin{abstract}
An alternative, pedagogically simpler derivation of the allowed physical wave
 fronts of a propagating electromagnetic signal is presented using geometric
 algebra.  Maxwell's equations can be expressed in a single multivector
 equation using 3D Clifford algebra (isomorphic to Pauli algebra spinorial
 formulation of electromagnetism).  Subsequently one can more easily solve
 for the time evolution of both the electric and magnetic field simultaneously
 in terms of the fields evaluated only on a 3D hypersurface.  The form of the
 special "characteristic" surfaces for which the time derivative of the
 fields can be singular are quickly deduced with little effort.
\keywords{} characteristics -- multivector -- clifford -- electromagnetism
\end{abstract}

\section{Introduction}
\quad Maxwell's equations (in a vacuum) are a set of (coupled) hyperbolic
 first-order differential equations.  As such, there are "characteristic"
 hypersurfaces over which the first derivatives of the fields can be
 discontinuous[1].  Specifically these are unique three-dimensional surfaces
 embedded in the four-dimensional continuum.  They correspond physically to
 the allowable physical wave fronts of a propagating electromagnetic
 disturbance of the field.

A derivation is provided by Adler[2] (based upon the earlier work of Fock[3])
 which shows the characteristics to be light-cones in Minkowski space;
 equivalently three-dimensional spheres expanding at the speed of light.
As the goal is to describe a three-surface, the problem is most naturally
 formulated in standard three-dimensional Gibbs vectors.  Hence the standard
 derivation consists of a rather lengthy manipulation of Maxwell's equations
 in vector form.  To arrive at the desired result requires many convoluted
 steps which can only be motivated by considerable experience in vector
 identities.

What is presented in this paper is an alternative derivation which is
 pedagogically simpler.  Maxwell's four equations are expressed in a single
 multivector equation using three-dimensional Clifford Algebra (isomorphic
 to the "spinorial formulation" of electromagnetism).  The notational
 economy of "four equations in one" cuts the number of steps by a
 corresponding factor of four.  Further the associativity of the Clifford
 product (replacing non-associative Gibbs cross product) and its duplicity
 (decomposes into inner and outer portions) simplifies the vector identities
 to transparency.  The derivation of equation for characteristics requires
 only two straight-forward steps.

The following section reviews the geometric algebra notation used in the
 derivation.  In section 3 the multivector formulation of Maxwell's equations
 is reviewed.  Section 4 contains the new Clifford algebra derivation of the
 characteristics for electromagnetic signals.

\section{Algebraic Notation}
\quad For the three-dimensional geometry the "Clifford" algebra[4,5,6] is
 isomorphic to the familiar Pauli matrix algebra ${\bf C}(2)$ [2-by-2
 complex matrices].  However, in contrast to standard view, the intrinsic
 8 degrees of freedom are given concrete geometric interpretation.

\subsection{The Clifford Group}

\quad The geometric interpretation of the multiplication rule,
$$\{ \sigma_j, \sigma_k\} = 2 \delta_{jk} , \quad (j,k=1,2,3) ,  \eqno(1)$$
is that perpendicular basis vectors anticommute.  The basis trivector:
 $i=\sigma_1 \sigma_2 \sigma_3 $ is associated with the unit volume
 (pseudoscalar).  As it commutes with all elements and has negative signature,
 Hestenes[4,6] declares it a geometric definition of the usual abstract $i$.
 Bivectors (pseudovectors) are direct products of 2 basis vectors, and
 geometrically associated with planes.  Multiplication of a vector by $i$
 yields the (Hodge) dual plane and visa versa, e.g.
$\sigma_1\sigma_2=i\sigma_3$.  The unit scalar $\{1\}$, 3 basis vectors
$\{ \sigma_j\}$, 3 basis bivectors $\{ i\sigma_j \}$ and trivector
$\{i\}$ make up the 8 element Clifford group of 3D orthogonal space.

\subsection{Vector Algebra}
\quad The direct "Clifford" product of two vectors can be decomposed into
  Grassmann symmetric "inner" (dot) and antisymmetric "exterior" (wedge)
 products respectively,
$${\bf AB} = {\bf A} \bullet {\bf B} + {\bf A}\wedge {\bf B}  ,\eqno(2)$$
where,
$$	{\bf A} \bullet {\bf B} = {\bf B} \bullet {\bf A} =
 \frac{1}{2}\{ {\bf A}, {\bf B} \}  = A_jB_k \delta^{jk},\eqno(3a)$$
$$	{\bf A} \wedge {\bf B} = -{\bf B} \wedge {\bf A} =
 \frac{1}{2} [ {\bf A}, {\bf B} ]  = i {\bf A}\times {\bf B}.\eqno(3b)$$
The symmetric part is the usual Gibbs dot product, and the wedge is a
 bivector, which in 3D is dual to the Gibbs cross product.  So eq. (2)
 is a {\it multivector}, i.e. a conglomorate of scalar plus bivector.

\subsection{Multivector Algebra}
\quad A general {\it multivector} or {\it cliffor}[5] is an aggregate
 sum of the four ranks of geometry with 8 degrees of freedom encoded,
$$ {\it M} = s + {\bf E }+ i {\bf H}+ i p,  \eqno(4)$$
where $s$ and $p$ are real scalars, and ${\bf E}$ and ${\bf H}$ are
 real vectors.
Under a parity inversion the odd ranked geometries will invert
 (vector ${\bf E}$, trivector $ip$), while the even (scalar $s$, bivector
  $i{\bf H}$ ) will not.  Hence in 3D we associate the alternate terms of
 {\it pseudovector} for bivector, and {\it pseudoscalar} for the trivector.

The geometric product of general multivectors is completely associative,
 $( {\it AB}) {\it C}= {\it A}( {\it BC})$.  The product of a vector with
 a bivector (or trivector) can also be decomposed into symmetric and
 antisymmetric portions.  We write down a summary of products and their
 Gibbs vector equivalents,
$${\bf A}\wedge {\bf B}\wedge {\bf C}= i[ {\bf A}\bullet(
 {\bf B}\times {\bf C})], \eqno(5a)$$
$${\bf A}\bullet( {\bf B}\wedge {\bf C})= - {\bf A}\times( {\bf B}\times
 {\bf C})=( {\bf A}\bullet {\bf B}) {\bf C}- ( {\bf A}\bullet {\bf C})
 {\bf B}, \eqno(5b)$$
$${\bf A}\bullet( {\bf B}\wedge {\bf C} \wedge {\bf D}) =
({\bf A}\bullet{\bf B}) {\bf C} \wedge {\bf D}-
({\bf A}\bullet{\bf C}) {\bf B} \wedge {\bf D}+
({\bf A}\bullet{\bf D}) {\bf B} \wedge {\bf C}. \eqno(5c)$$

\section{Multivector Electromagnetism}
\quad Maxwell's four equations can be expressed in a single multivector
 form using the geometric algebra.  It is this notational economy that will
 allow us a simple derivation of the characteristics.  Here we review the
 Clifford algebra formulation of electrodynamics.

\subsection{Vector Calculus}
\quad The application of a gradient operator$^4$ on a vector field yields
 a scalar plus bivector (pseudovector) field,
$$\nabla {\bf E} = \nabla \bullet {\bf E} + \nabla \wedge{\bf E}=
 \nabla \bullet {\bf E} + i(\nabla \times{\bf E}), \eqno(6)$$
Note the interpretation differs from standard in that $( \nabla \times
 {\bf E})$ is a vector (not a pseudovector!).  Multipled by $i$, it becomes
 a bivector/pseudovector.

\subsection{Multivector Field}
\quad We define the electromagnetic field multivector[5,6,7] to be the
 aggregate sum of "vector" electric plus "bivector" (pseudovector) magnetic
 fields, $F = {\bf E} + i{\bf H}$.  The gradient of the field has all four
 ranks of geometry represented,
$$\nabla F = (\nabla\bullet{\bf E}) -(\nabla\times{\bf H})+
i (\nabla\times{\bf E}) +i(\nabla\bullet{\bf H}),\eqno(7)$$
in the order scalar, vector, bivector (pseudovector) and trivector
 (pseudoscalar) respectively (parenthesis included for brevity).

\subsection{Maxwell's Equations}
\quad One can encode all four Maxwell's equations in the single
 multivector equation[5,6],
$$c \nabla F +  \partial_t F = c\rho -  {\bf J}	,\eqno(8a)$$
(assuming Heaviside-Lorentz units where $c$ is the speed of light),
 which in the sourceless case becomes,
$$	c \nabla F = -\partial_tF 	.\eqno(8b)$$
Each of the 4 distinct geometric parts of eq. (8a) yields one of the
 standard Maxwell equations.  Specifically, the scalar part of eq. (8a)
 yields Gauss's law, the vector portion is Ampere's law, the bivector is
 Faraday's law and the trivector is the magnetic monopole equation.

\section{The Characteristic Solution}
The goal is to obtain an equation for the time evolution of the multivector
 field $F$ in terms of the field evaluated only on the the hypersurface.
 The presentation here will parallel that of Adler[2] (which is based on
 the earlier work of Fock[3]), except in the economical multivector
formulation.

\subsection{Hypersurface Description}
\quad The equation of a smooth three-dimensional hypersurface {\it S}
 embedded in the four-dimensional space-time manifold is parameterized,
$$\omega(x^0,x^1,x^2,x^3) = h({\bf r}) - x^0 = 0	,	\eqno(9)$$
where $\omega(x^0,x^1,x^2,x^3)$ is continuous in first order derivatives,
 $x^o= ct$ and $\partial\omega/\partial x^0$ must be nonzero [so that one
 can construct function $h({\bf r})$ ].  The multivector field
$\hat F=F({\bf r}) , {\bf r} )$, on surface {\it S} is a function only of the
 space coordinates ${\bf r }=(x,y,z)$, where we have adopted the "hat"
 notation of Adler[2].

\subsection{Fock Relations}
Assuming the vector functions $\hat {\bf E}({\bf r}) = {\bf E}( h( {\bf r}),
 {\bf r})$ and $\hat {\bf H}( {\bf r})= {\bf H}( h( {\bf r}), {\bf r})$
 have continuous first derivatives, the chain rule provides,
$$\partial_j \hat F = \partial_jF  +
 \frac{1}{c}(\partial_tF \   \partial_jh)	.\eqno(10a)$$
Contracting with the basis vectors $\sigma_j$ yields the multivector form
 of the relations stated by Fock (i.e. contains equations 3.06-3.09 of ref[3]
),
$$\nabla \hat F = \nabla F  +  \frac{1}{c} \nabla h
 ( \partial_t F )	.\eqno(10b)$$
Substituting the multivector form eq. (8b) of Maxwell's equations for
 $\nabla F$,
$$\nabla \hat F = \frac{1}{c}(\nabla h - 1)  ( \partial_t F )	,\eqno(10c)$$
gives a single multivector equation containing four identities derived by
 Adler (equations 4.22-4.25 of ref[2] respectively).

\subsection{Characteristic Equation}
\quad The problem is now simply to invert eq. (10c) to obtain an equation for
 $\partial_t F $ in terms of only the field evaluated on the the hypersurface,
 i.e. in terms of $\hat F$ (and its space derivatives).  The derivation is as
 trivial in the multivector formulation as it is tedious in standard vector
 analysis.  We simply multiply eq. (10c) on the left by the multivector
 $(\nabla h + 1)$ to "scalarize" the right side,
$$(\nabla h + 1) \nabla \hat F = \frac{1}{c}(|\nabla h|^2 - 1)
 ( \partial_t F )	.\eqno(11)$$
In principle we are done; we need not look at each geometric component to
 deduce the nature of the characteristics.

If the scalar factor of $(|\nabla h|^2 - 1)$ which appears on the right side
 of eq. (11) is zero, then the time derivative of the field could be
 discontinuous across {\it S}.  From this point on the discussion is the same
 as given in the references[2,3].  The equation that these "characteristic"
 hypersurfaces must obey is determined by the condition,
$$|\nabla h |^2 = 1	,\eqno(12a)$$
which according to the definition of the hypersurface is equivalent to ,
$$(\partial_t\omega)^2 - (\nabla \omega )^2 = 0. \eqno(12b)$$
The propagation of an electromagnetic wave front must satisfy this equation.
 Some particular hypersurface solutions are of the form,
$$\omega(x^0,x^1,x^2,x^3) = |{\bf r} -{\bf r}_0 | - c (t- t_0), \eqno(13a)$$
$$\omega(x^0,x^1,x^2,x^3) = {\bf n} \bullet {\bf r} - ct, \eqno(13b)$$
corresponding to a spherical wave front (expanding at speed of light about
 point ${\bf r}_0$), and  a plane wave front (moving at the speed of light,
 in direction of unit vector  ${\bf n} $) respectively.

\section{Summary}
\quad Deriving the characteristics of Maxwell's equations via Clifford
 algebra formulation is embarrassingly trivial, essentially consisting of
 eqs. (10bc) and (11).  The notation encodes four-equations-in-one such that,
 for example, Maxwell equations, the Fock relations, and Adler's identities
 are each represented by a single compact multivector statement [equations
 (8a), (10b) and (10c) respectively].  It remains to be seen if geometric
 algebra would provide similar clarity in the derivation of characteristics
 in other areas, n.b. electromagnetic waves in media, mechanical shock waves
 in solids or fluids, the Proca equation, relativistic quantum mechanics and
 gravitation.

\section{Acknowledgements}
	The author expresses his appreciation to Professors J. Lawrynowicz,
 Institute of Mathematics, Polish Academy of Sciences, and J. Rembielinski,
 Institute of Physics, University of Lodz, for the invitation to visit,
 arranging for financial support and the hospitality accorded him during
 his stay in Poland.

\end{document}